\documentclass[12pt]{iopart}

\usepackage{amsfonts,amssymb,bm}
\usepackage{cancel}
\usepackage{graphicx}
\usepackage{mathcomp}
\usepackage{color}
\usepackage{wrapfig}
\usepackage[english]{babel}

\usepackage[bookmarksopen=true, pdftitle={Notes}, pdfauthor={Sara Gallian}, colorlinks=true, linkcolor=black, citecolor=black, pdfstartview=FitH]{hyperref}

\definecolor{rubblau}{cmyk}{1.0,0.5,0.0,0.6}
\definecolor{rubgruen}{RGB}{141,174,16}
\definecolor{hellgrau}{rgb}{0.8,0.8,0.8}
\definecolor{bord}{RGB}{128,0,0}
\definecolor{jade}{RGB}{0,168,107}

\newcommand{\eps}{\varepsilon}	        
\newcommand{\dd}{\partial}

\begin{document}

\title[Phenomenological model for description of rotating spokes in HiPIMS discharges]{A phenomenological model for the description of rotating spokes in HiPIMS discharges}

\author{S Gallian$^1$, W N G Hitchon$^2$, D Eremin$^1$, T Mussenbrock$^1$, R P Brinkmann$^1$}

\address{$^1$ Institute for Theoretical Electrical Engineering, Ruhr University Bochum, D-44780 Bochum, Germany}
\address{$^2$ Department of Electrical and Computer Engineering, University of Wisconsin-Madison, WI 53706 Madison, Wisconsin, USA}

\ead{gallian@tet.rub.de}

\date{\today}

\begin{abstract}
In the ionization region above circular planar magnetrons, well defined regions of high emissivity are observed, when the discharge is driven in the HiPIMS regime. 
These regions are characterized by high plasma density and are often referred to as ``spokes".
Once their mode is stabilized, these structures rotate in the $\textbf{E} \times \textbf{B}$ direction with a constant rotation frequency in the hundreds of kHz range. A phenomenological model of the phenomenon is developed, in the form of a system of nonlinear coupled partial differential equations. The system is solved analytically in a frame co-moving with the structure, and its solution gives the neutral density and the plasma density once the electron density shape is imposed. 
From the balance of ionization, electron loss and constant neutral refilling, a steady state configuration in the rotating frame is achieved for the electron and neutral densities. Therefore, the spoke experimentally observed can be sustained simply by the combination of this highly reduced number of phenomena.
Finally, a study of the sensitivity of neutral and plasma densities to the physical parameters is also given.

\end{abstract}

\maketitle
\section{Introduction}
High Power Impulse Magnetron Sputtering (HiPIMS) \cite{1999-Kouznetsov} is a recently developed Ionized Physical Vapor Deposition (IPVD) technique, able to achieve an ultra dense plasma with a high ionization degree among the sputtered atoms. These characteristics are particularly desirable since they allow film growth control and ensure high film quality (see e.g. \cite{2012-GudmundssonBrenning},\cite{2010-SarakinosAlami} and references therein). 
In a HiPIMS system a very high peak power density (several kW/cm$^{2}$) is delivered to a conventional magnetron in a pulsed fashion: a large bias is applied to the target in short pulses of duration of a few hundred microseconds with a low duty cycle ($0.5$-$5 \%$). \\
In the plasma region near the target, the magnetized electrons describe a spiral-like gyromotion around a field line and exhibit a bouncing motion along the magnetic field lines. They are also subject to a series of additive drifts (the \textbf{E} $\times$ \textbf{B} drift, and the drifts due to the magnetic field gradient and curvature) which are all in the azimuthal direction in the plane of the target. 
According to the definition in \cite{2012-AndersNi}, electrons exhibit a ``closed drift" and are bound to remain in the near target region, allowing for a large amount of neutral ionization to occur, even when the collision mean free path is comparable to the system dimensions. \\
Given the geometry of the set up, a circular planar magnetron, and the closed electron drift, it would be reasonable to assume azimuthally symmetric macroscopic discharge properties. Indeed, the discharge appears homogeneous to the naked eye, it is only when captured with a short exposure camera (acquisition time less than or equal to 100 ns) that high emissivity regions breaking the axial symmetry are noted. The presence of rotating structures has been reported independently by different groups: Kozyrev \emph{et al.} \cite{2011-KozyrevSochugov}, Ehiasarian \emph{et al.}\cite{2012-EhiasarianHecimovic} and Anders \emph{et al.} \cite{2012-AndersNi}. 
According to the experimental observations, these well defined regions of high emissivity rotate with constant frequency of the order of $50$-$100$ kHz. 
Moreover, in \cite{2012-AndersNi} it is clearly shown that these emissivity peaks are associated with electron flares away from the target, which may be responsible for the observed enhanced electron transport across the field lines in HiPIMS. \\
Since the plasma is mostly concentrated in these regions, it is reasonable to believe that the observed rotating spoke may determine the average plasma density, carry most of discharge current and influence the cross field electron transport. Therefore understanding the formation, evolution and behavior of these regions plays a fundamental role in characterizing the overall discharge.\\
Anders \emph{et al.} in \cite{2012-AndersNi, 2012-Anders-SelfOrganization} also attempted to explain the formation and evolution of the structures as a balance between a positive feedback process, related to regions of higher ``stopping power" for electrons in closed drift, and self-organization mechanisms, due to a difference between the zone transit time and the time sputtered atoms need to reach the ionization region.\\
More recently, Brenning \emph{et al} in \cite{2013-BrenningLundin} proposed a ``unified spoke model" describing spokes both in HiPIMS and critical ionization velocity (CIV) discharges. In particular, an electron heating mechanism that combines secondary electron emission and wave-particle interaction is proposed and it is theorized to have a threshold in proximity of the Alfv\'{e}n CIV: the spoke angular velocity would therefore be regulated.\\
This contribution has the aim of providing a phenomenological model for a single rotating spoke.  
To devise such a model, we start by considering the evolution of a set of chemical species, subject to drift and diffusion processes. Then we simplify the resulting system of advection, diffusion and reaction equations retaining only the terms that we believe responsible for sustaining the single structure considered. We then comment on the correspondence of this simplified model to experimental observations.


\section{Phenomenological system set up} \label{syssetup}
The differential form of the continuity equation for the species $s$ reads
\begin{equation}
\frac{\dd n_{s}(\mathbf{x},t)}{\dd t} + \nabla \cdot \mathbf{\Gamma}_{s}(\mathbf{x},t) = R_{s}(\mathbf{x},t), \qquad s = 1, ..., N_{s} \label{ADR}
\end{equation}
where $n_{s}$ is the density of species $s$, $\mathbf{\Gamma}_{s}$ is the total flux, and $R_{s}$ represents the net generation of $n_{s}$ and accounts for both source and sink terms, including the interaction of species $s$ with all other species considered. 
The flux $\mathbf{\Gamma}_{s}$ is considered to have both an advective and diffusive contribution, so that $\mathbf{\Gamma}_{s} = \mathbf{u_{s}} \ n_{s} - D_{s} \nabla n_{s}$, where $\mathbf{u}_{s}$ is the drift velocity and $D_s$ the diffusion coefficient. \\
The partial differential equation \eref{ADR} is an Advection-Diffusion-Reaction equation which generally models the evolution of a chemical species that undergoes reaction, can diffuse in the solvent and is transported. Here, it will be employed to model the behavior of electrons and neutrals.  \\
Consider a single emission structure or spoke during a HiPIMS discharge, represented as a peak in the electron density, which is rotating with constant angular velocity $\Omega$. 
The electrons move in a closed drift, dominated by the \textbf{E} $\times$ \textbf{B} velocity, but are observed to be slowed down: the spoke region rotates with a speed of about 10 \% of the \textbf{E} $\times$ \textbf{B} drift \cite{2012-AndersNi}.
Therefore, the electrons are considered to move with a collective drift velocity proportional to $\Omega$ and are free to diffuse from the higher density region (i.e. the spoke region) to the lower density ones. The electron density evolution is described by an advection-diffusion-reaction equation 
\begin{equation}
	\frac{\dd n_{\rm e}}{\dd t} + \Omega \frac{\dd n_{\rm e}}{\dd \theta} - D_\theta \frac{\dd^2 n_{\rm e}}{\dd \theta^2}  = S_e - L_e, \label{neADR}
\end{equation}
where $D_\theta = D/r^2$ is an effective angular diffusion coefficient in the azimuthal direction, and $S_e$ and $L_e$ are source and loss terms for the electron density.
Since the electron plasma frequency $\omega_{\rm pe}$ is much larger than all characteristic frequencies of the system, quasi-neutrality holds outside the thin sheath and can safely be assumed during the discharge.\\
On the other hand, the neutral species, argon and sputtered metal, are subject to drift and diffusion on a much longer time scale than the electron time scale and the structure rotation period. Therefore, the rate equations for argon and sputtered metal reduce to reaction equations
\begin{equation*}
	\frac{\dd n_{j}}{\dd t}  = S_{j} - L_{j}, \quad j \mbox{ = Ar, M}.
\end{equation*}
Following \cite{2011-RaaduAxnaes} the source and loss terms for the neutral species can be specified as follows
\numparts
\begin{eqnarray}
\fl \eqalign{\frac{\dd n_{\rm Ar}}{\dd t} =& -k_{\rm iz,Ar} n_{\rm e} n_{\rm Ar} + k_{\rm P} n_{\rm Ar*} n_{\rm M} 
									   + k_{\rm chexc} n_{\rm Ar+} n_{\rm M} 
									   - k_{\rm exc} n_{\rm e} n_{\rm Ar} 
									    + k_{\rm dexc} n_{\rm e} n_{\rm Ar*} \\&+ \frac{\Gamma_{\rm Ar}}{L} - L_{\rm wind},}\\
\fl \eqalign{	\frac{\dd n_{\rm M}}{\dd t} =& -k_{\rm iz,M} n_{\rm e} n_{\rm M} - k_{\rm P} n_{\rm Ar*} n_{\rm M} 
									   - k_{\rm chexc} n_{\rm Ar+} n_{\rm M} -
									    \frac{\Gamma_{\rm M}}{L} \\& + (\Gamma_{\rm Ar+} Y_{\rm sputt}+ \Gamma_{\rm M+} Y_{\rm selfsputt}) \frac{S_{\rm RT}}{V}.}	\label{nArnM}
\end{eqnarray}
\endnumparts
In both equations the first three terms account for ionization by electron collisions, Penning ionization and charge exchange collisions respectively. $k_{\rm exc} n_{\rm e} n_{\rm Ar}$ and $k_{\rm dexc} n_{\rm e} n_{\rm Ar*}$ describe excitation and de-excitation of the metastable level $n_{\rm Ar*}$. $\Gamma_{\rm Ar}/L$ is the refill rate of argon gas from outside the ionization region, which depends on the discharge pressure. The last term in the Ar rate equation represents the loss of gas due to the metal sputter wind: argon is physically pushed away from the region near the target by collisions with the highly energetic sputtered metal (see i.e. \cite{1998-Rossnagel-IPVD} and \cite{2010-HorwatAnders}). In the rate equation for the metal, ${\Gamma_{\rm M}}/{L}$ is the loss rate by diffusion outside the region of interest, and finally the last term represents the source of metal ions due to sputtering or self-sputtering of the target determined by the fluxes of argon and metal ions multiplied by the respective sputter yields. Finally, the geometry parameters $S_{\rm RT}$, $V$ and $L$ are the racetrack surface area, the ionization region volume and the extent of the ionization region in the direction perpendicular to the target ($z$-direction).\\
With the aim of devising a simplified model for the rotating structure, the set of equations is further reduced  to account for a single neutral species density: $n_{\rm n}(\theta,t) = n_{\rm Ar}+n_{\rm M}$ and $n_{\rm M} \equiv \gamma_{\rm M} \ n_{\rm n}$, where $\gamma_{\rm M}$ is the fraction of metal neutrals with respect to the total neutral density. The sum of the two rate equations in \eref{nArnM} reads
\begin{equation}
\fl \eqalign{\frac{\dd n_{\rm n}}{\dd t} =& - \left( k_{\rm iz,Ar} n_{\rm Ar} + k_{\rm iz,M} n_{\rm M}\right) n_{\rm e}+ 
							    \left(- k_{\rm exc} n_{\rm Ar} +  k_{\rm dexc}  n_{\rm Ar*} \right) n_{\rm e} +							
	\frac{\Gamma_{\rm Ar}}{L} 
	- L_{\rm wind} - \frac{\Gamma_{\rm M}}{L} \\&+ (\Gamma_{\rm Ar+} Y_{\rm sputt}+ \Gamma_{\rm M+} Y_{\rm selfsputt}) \frac{S_{\rm RT}}{V} \\
		& \approx - \left[ k_{\rm iz,Ar} (1-\gamma_{\rm M}) + k_{\rm iz,M} \gamma_{\rm M}\right] n_{\rm n} n_{\rm e} +
		\frac{\Gamma_{\rm Ar}}{L} - \frac{\Gamma_{\rm M}}{L} + \\& \Gamma_{\rm e}[(1-\gamma_{\rm M+}) Y_{\rm sputt}+ \gamma_{\rm M+} Y_{\rm selfsputt}] \frac{S_{\rm RT}}{V}, \label{nnSteps}
}
\end{equation}
where the hypothesis of quasi-neutrality ($n_{\rm M+}+ n_{\rm Ar+} =  n_{\rm e}$) has been used together with the definition of the ratio $\gamma_{\rm M+}=n_{\rm M+}/n_{\rm e}$ to rewrite the sputtering term. \\
Penning ionization and charge exchange terms simplify in the summation. 
Excitation and de-excitation events have been neglected as well as the loss of argon gas due to the sputtering wind. The latter term has been reported \cite{2012-HuoRaadu} to have a smaller influence than the Ar depletion by ionization at low pressures. Moreover, even though neglecting excitation processes is a strong and likely over-simplifying assumption, global modeling efforts by Raadu \emph{et al.} in \cite{2011-RaaduAxnaes} show that during the main pulse ($15\ \mu $s to $100\ \mu$s), following the ignition and breakdown phase, the plasma density varies mostly because of electron impact ionization and of ion loss fluxes across the system boundaries. Indeed a crude estimate of the excitation and de-excitation term is performed in \ref{Appendix} and the rate is found to be negligible with respect to the ionization term.  \\
When source and loss terms in equation \eref{neADR} are specified consistently with \eref{nnSteps}, the system to be solved reads
\numparts
\begin{eqnarray}
	\frac{\dd n_{\rm e}}{\dd t} + \Omega \frac{\dd n_{\rm e}}{\dd \theta} -  D_\theta \frac{\dd^2 n_{\rm e}}{\dd \theta^2}  &= \bar{k}_{\rm ion} n_{\rm e} n_{\rm n}- \nu_{\rm l}n_{\rm e}, \label{neLF}\\
	\frac{\dd n_{\rm n}}{\dd t}  &= - \bar{k}_{\rm ion} n_{\rm e} n_{\rm n}+ R, \label{nnLF}
\end{eqnarray}
\endnumparts
where $\bar{k}_{\rm ion} =  k_{\rm iz,Ar} (1-\gamma_{\rm M}) + k_{\rm iz,M} \gamma_{\rm M} \ $[m$^3$/s] is the ionization rate coefficient, $\nu_{\rm l}\ $[Hz] is the electron loss frequency across the magnetic field lines and 
\begin{equation*}
R = ({\Gamma_{\rm Ar}}-\Gamma_{\rm M})/{L} + \Gamma_{\rm e,z}[(1-\gamma_{\rm M+}) Y_{\rm sputt}+ \gamma_{\rm M+} Y_{\rm selfsputt}] {S_{\rm RT}}/{V}, \quad {\rm [m^{-3}/s] }
\end{equation*}
accounts for the net source of neutrals in the domain through sputtering and diffusion. Finally, to enforce symmetry, periodic boundary conditions have to be satisfied: $n_{\rm e}(0,t) = n_{\rm e}(2\pi,t)$ and $n_{\rm n}(0,t) = n_{\rm n}(2\pi,t)$, likewise the derivatives have to be matched at the domain boundaries.\\
The ionization rate $\bar{k}_{\rm ion}$ is a functional of the electron energy
distribution. In the case of a Maxwellian, it is solely a function of the electron temperature and independent of the electron density. 
For the HiPIMS regime, however, there is experimental evidence that the plasma becomes more active (or ``hotter") with increasing density. Therefore, in the absence of a separate energy balance this effect is modeled by employing the linear ansatz
\begin{equation*}
	\bar{k}_{\rm ion}(n_{\rm e}) = k_{\rm ion} + \beta n_{\rm e}.
\end{equation*}
Therefore, equations \eref{nnLF} and \eref{neLF} are rewritten as
\numparts
\begin{eqnarray}
	\frac{\dd n_{\rm e}}{\dd t} + \Omega \frac{\dd n_{\rm e}}{\dd \theta} -  D_\theta \frac{\dd^2 n_{\rm e}}{\dd \theta^2}  &= k_{\rm ion} n_{\rm e} n_{\rm n} + \beta n^2_{\rm e} n_{\rm n} - \nu_{\rm l}n_{\rm e}, \label{neLF2}\\
	\frac{\dd n_{\rm n}}{\dd t}  &= - k_{\rm ion} n_{\rm e} n_{\rm n} - \beta n^2_{\rm e} n_{\rm n}+ R. \label{nnLF2}
\end{eqnarray}
\endnumparts
It is assumed that this simplified system is able to phenomenologically reproduce the rather complicated ionization structure, after the initial transient and moving at a constant angular velocity $\Omega$ in the azimuthal direction. 

\subsection{Coordinate transformation in rotating frame}
Introducing the non-dimensional quantities: $ T = t \ \Omega$ and $ n'_j = \frac{\Omega}{R} n_j \  (j$ = e,n), the non-dimensional system reads
\numparts
\begin{eqnarray}
	\frac{\dd n'_{\rm e}}{\dd T} +  \frac{\dd n'_{\rm e}}{\dd \theta} -D'_\theta \frac{\dd^2 n'_{\rm e}}{\dd \theta^2}  &= k'_{\rm ion} n'_{\rm e} n'_n +\beta' n_{\rm e}^{'2} n'_n- \nu'_{\rm l}n'_{\rm e}, \label{neLFNorm}\\
	\frac{\dd n'_n}{\dd T}  &= -  k'_{\rm ion} n'_{\rm e} n'_n - \beta' n_{\rm e}^{'2} n'_n+ 1. \label{nnLFNorm}
\end{eqnarray}
\endnumparts
where the non-dimensional physical parameters are defined as: 
$k'_{\rm ion} = k_{\rm ion} R/\Omega^2$, $\nu'_{\rm l} = \nu_{\rm l}/\Omega$, $D'_\theta = D_\theta/\Omega$.\\
It is here convenient to consider a reference frame that is rotating with constant angular speed $\Omega$, defined by the transformation
\begin{equation*}
	n'_{\rm e}(\theta,T) = n'_{\rm e}(\hat{\theta}(\eta,\tau),\tau) = \hat{n}_{\rm e}(\eta,\tau), \quad \mbox{with }
	\cases{
		 \hat{\theta} =  \eta + \tau \\
		T = \tau
	}.
\end{equation*}
Dropping the apexes and the circumflexes for convenience, the non-dimensional system of rate equations expressed in the moving frame $(\eta,\tau)$ reads
\numparts
\begin{eqnarray}
\label{SysMFNorm-TV}
		\frac{\dd n_{\rm e}}{\dd \tau}-  D_\theta \frac{\dd^2 n_{\rm e}}{\dd \eta^2}  = k_{\rm ion} n_{\rm e} n_{\rm n}  +  \beta {n}^2_{\rm e} n_{\rm n} -\nu_{\rm l}n_{\rm e}, \\
		\frac{\dd n_{\rm n}}{\dd \tau}  - \frac{\dd n_{\rm n}}{\dd \eta}= -k_{\rm ion} n_{\rm e} n_{\rm n} - \beta {n}^2_e n_{\rm n} + 1 .
\end{eqnarray}
\endnumparts
Finally, experimental frequency domain analysis shows a transition from stochastic behavior to periodic behavior of the system, when a single spoke remains \cite{2013-WinterHecimovic}. The single peak rotating with constant velocity $\Omega$ is thus considered a steady state configuration.
Therefore, we look for a stationary solution in the co-moving reference frame. The system \eref{SysMFNorm-TV} reduces to
\numparts
\begin{eqnarray}
\label{SysMFNorm}
	-  D_\theta \frac{\dd^2 n_{\rm e}}{\dd \eta^2}  = k_{\rm ion} n_{\rm e} n_{\rm n}  +  \beta {n}^2_e n_{\rm n} -\nu_{\rm l}n_{\rm e}, \\
	 - \frac{\dd n_{\rm n}}{\dd \eta}= -k_{\rm ion} n_{\rm e} n_{\rm n} - \beta {n}^2_e n_{\rm n} + 1 .
\end{eqnarray}
\endnumparts


\subsection{System solution}
\begin{figure}[h!] 
\centering  {\includegraphics[width=8cm]{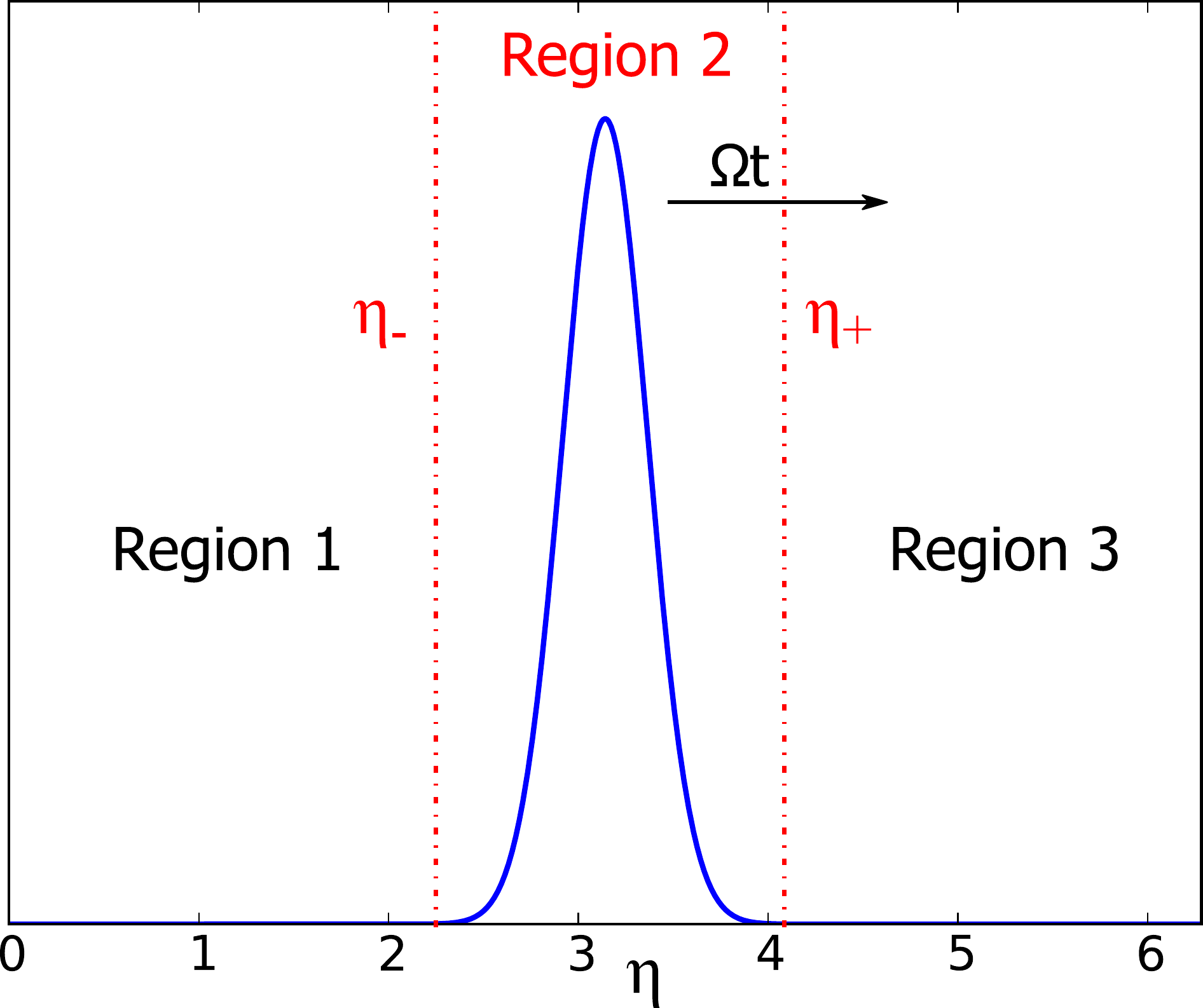}} \caption{Gaussian function chosen as shape for the electron density.}\label{RegionGaussian}
\end{figure}

The first equation of the system \eref{SysMFNorm} is a nonlinear eigenvalue problem for $n_{\rm e}(\eta)$, so that its profile needs to be assumed. 
To phenomenologically reproduce the spoke, the electron profile density is assumed to be a single or a sum of Gaussian profiles rotating with constant speed $\Omega$. For simplicity here a single Gaussian is imposed
\begin{equation}
	n_{\rm eg}(\eta) = \frac{A}{\sqrt{2 \pi W}} \exp \left ({-\frac{(\eta - \eta_0)^2}{2 W}} \right).   \label{gaussian}
\end{equation}
The coefficients $A$ and $W$ are treated as free parameters and can in general be functions of time in the laboratory frame. \\
Even though there are indeed multiple ways of constructing a density profile to describe a spoke-like structure, the Gaussian shape has been chosen because of its resemblance to the experimental observation: a Gaussian tends to zero smoothly and over a wide region.
The profile imposed for the electron density has the aim of describing a narrow rotating perturbation: the domain $(0,2 \pi)$ can be divided into three intervals, as sketched in \fref{RegionGaussian}.\\
In regions 1 and 3  $n_{\rm eg}(\eta)$ is approximately 0, as opposed to region 2. Thus the solution for the neutral density can be found separately in these three intervals:  in the first and the third regions only the source term is taken into account, while in the central region, where the spoke is located, only the ionization terms are considered.
After the analytic solution has been found for the neutral density in region 2, it is substituted in the equation for the electron density \eref{SysMFNorm}, which is averaged between $\eta_{\rm -}$ and $\eta_{\rm +}$, and solved for the amplitude A.

Region 2 is delimited by $\eta_{\rm +}$ and $\eta_{\rm -}$, that are correlated to the width of the Gaussian by
\begin{equation*}
	\eta_\pm = \pi \pm k_{p} \sqrt{W} .
\end{equation*}
A reasonable value for $k_{p}$, that allows one to maintain the errors due to the division in intervals within tolerances, is found to be $3$. \\
Solving in the three regions separately allows us to obtain analytic solutions for the neutral density in each region. The solutions need to be matched at the interfaces $\eta_{\rm +}$ and $\eta_{\rm -}$ and periodicity has to be imposed. The neutral density throughout the domain which results is
\begin{equation*}
n_{\rm n}(\eta) = 
\cases{
	-\eta + \eta^- + n_{\rm n}(\eta^-), \quad \eta \in (0,\eta^-) \\
	n_{\rm n}(\eta^-) \exp \int_{\eta^-}^\eta \left( k_{\rm ion} n_{\rm eg} + \beta {n}^2_{eg} \right) d\eta, \quad \eta \in (\eta^-,\eta^+)\\
	-\eta + 2 \pi + \eta^-+n_{\rm n}(\eta^-), \quad \eta \in (\eta^+,2\pi)
}
\end{equation*}
where the integral $\int_{\eta^-}^\eta \left( k_{\rm ion} n_{\rm eg} + \beta {n}^2_{eg} \right) d\eta$ has a closed form for the choice of $n_{\rm e}$ and $n_{\rm n}(\eta^-)$ is determined when the solutions from the three regions are matched at the boundaries.
The neutral density in region 2 reads
\begin{equation*}
\fl \eqalign{n_{\rm n2} &= n_{\rm n}(\eta^-) \exp\left\{ \frac{1}{4} A \left[ 2 k_{\rm ion} \left(1- {\rm erf} \left(\frac{\pi-\eta}{\sqrt{2 W}}\right)\right) + \beta \frac{A}{\sqrt{\pi W}}\left(1- {\rm erf} \left(\frac{\pi-\eta}{\sqrt{W}}\right)\right) \right]\right\},\\
&{\rm where }\quad  n_{\rm n}(\eta^-) = 2 \ (\pi - k_p \sqrt{W}) \left[-1 + \exp\left( A k_{\rm ion} + \frac{A^2}{2 \sqrt{\pi W}}\beta \right)\right]^{-1}
.}
\end{equation*}
Here erf$(x)$ is the error function defined as: 
${\rm erf}(x) = \frac{2}{\sqrt{\pi}} \int_0^x e^{-t^2} dt$.
The amplitude coefficient $A$ in the electron density $n_{\rm eg}$ can be determined from the solution of the equation for the electron density in region 2
\begin{equation}
	-  D_\theta \frac{\dd^2 n_{\rm eg}}{\dd \eta^2} +\nu_{\rm l}n_{\rm eg}  = k_{\rm ion} n_{\rm eg} n_{\rm n}  +  \beta {n}^2_{eg} n_{\rm n} = \frac{\dd n_{\rm n2}}{\dd \eta}. \label{neSSMV}
\end{equation}
Taking the integral of \eref{neSSMV} between $\eta^-$ and $\eta^+$, and assuming the width $W$ of the perturbation to be small compared to the whole domain, the coefficient $A$ can be analytically determined as
\begin{equation}
	A = \frac{ -\eta^+ +  2 \pi + \eta^-}{\nu_{\rm l}}=\frac{2}{\nu_{\rm l}}\left(\pi-k_p \sqrt{W}\right) .
\end{equation} 
Given the normalization chosen for $n_{\rm eg}$, $A$ is approximately the maximum value of the electron density. In dimensional quantities the peak electron density is
\begin{equation}
	n_{\rm eg}(\eta_0) = \sqrt{2} \ \frac{\pi-k_p \sqrt{W}}{\sqrt{\pi W}} \frac{R}{\nu_{\rm l}}. \label{nepeak}
\end{equation} 
The result \eref{nepeak} relates the maximum density to the width ${W}$ of the electron density perturbation (i.e. the ionization region), the electron loss frequency $\nu_{\rm l}$ and the neutral refill rate $R$.
As will be argued in detail in \ref{Appendix}, $\nu_{\rm l}$ is determined by the electron drift and diffusion away from the target weighted by the time it takes the single structure to perform a full rotation. On the other hand, the width of the spoke structure depends on the electron diffusion in the azimuthal direction and the neutral ionization frequency. Finally, the neutral source rate defines the quantity of material that can be ionized. 
Therefore it is reasonable that the peak electron density is determined by the concurrence of these three terms.\\
The neutral density may be estimated as proportional to $R/\Omega$, i.e. to how many neutrals are generated during one period of rotation of the ionization structure. In particular, the peak neutral density, in dimensional quantities is
\begin{equation*}
	n_{\rm n}(\eta^+) = \xi_p \left[ 1 + {\rm coth} \left(R \xi_p \ \frac{  R \xi_p \beta + \nu_{\rm l}\sqrt{\pi W} k_{\rm ion}}{2 \pi \Omega \nu_{\rm l}^2 \sqrt{\pi W} }\right) \right] \frac{R}{2 \pi \Omega},
\end{equation*}
where $\xi_p = \pi - k_p \sqrt{W}$.
Neglecting the $\beta$ term, the peak density simplifies to
\begin{equation}
\fl	n_{\rm n}(\eta^+) = \xi_p \left[ 1 + {\rm coth} \left(\xi_p \ \frac{ R k_{\rm ion}}{2 \pi \Omega \nu_{\rm l}}\right) \right] \frac{R}{2 \pi \Omega} = \xi_p \left[ 1 + {\rm coth} \left(\xi_p \ \frac{ k'_{\rm ion}}{\nu'_{\rm l}}\right) \right] \frac{R}{2 \pi \Omega}, \label{nnpeaknndim}
\end{equation}
here $k'_{\rm ion}$ and $\nu'_{\rm l}$ are the non-dimensional coefficients. From considerations carried out in \ref{Appendix}, the values of $k'_{\rm ion}$ and $\nu'_{\rm l}$ have the same order of magnitude. The term in square brackets in \eref{nnpeaknndim} varies between 2 and 5 for $k'_{\rm ion}/\nu'_{\rm l}= 0.1 \div 10$ . Therefore, the neutral density is mainly determined by the ratio $R/\Omega$.

\section{Results and discussion} \label{results}
\begin{figure}[h!] 
\centering  {\includegraphics[width=8cm]{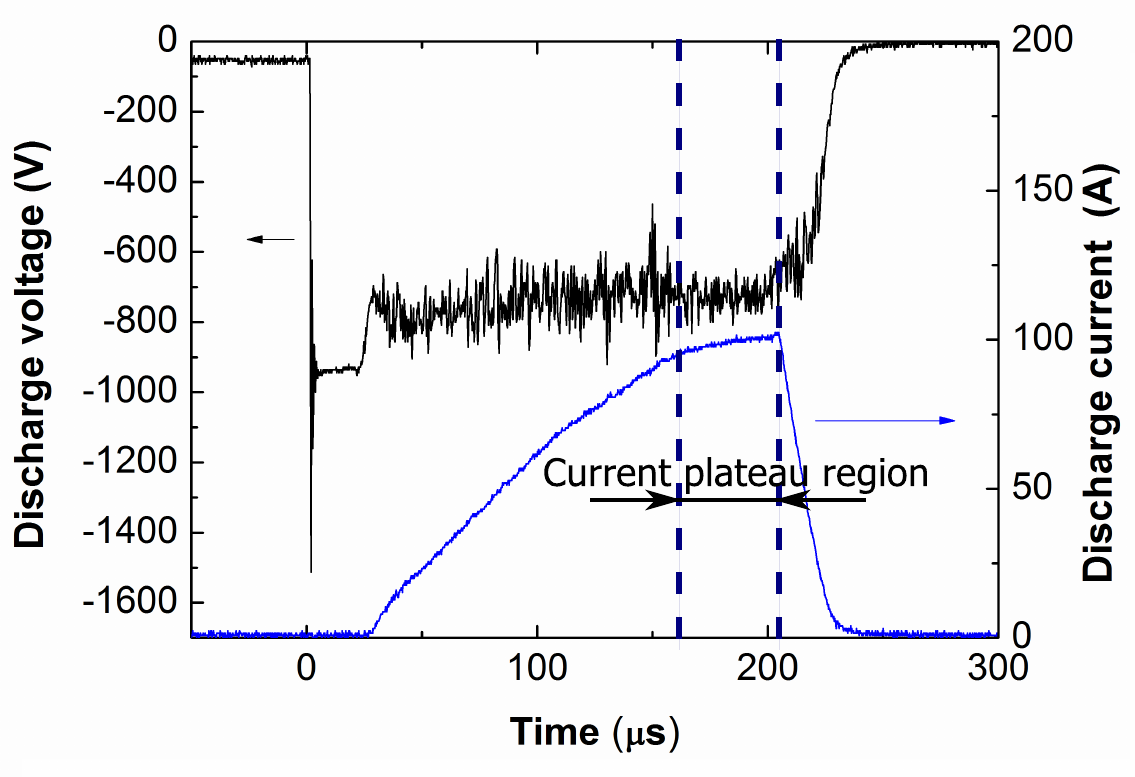}}\caption{Discharge voltage and current profile of the discharge taken from Hecimovic \emph{et al.} \cite{2013-HecimovicGallian} }\label{BiasVsCurrent}
\end{figure}

The geometrical parameters that define the analytic solution are specified to describe the experimental set up described in \cite{2012-EhiasarianHecimovic}. The physical parameters refer to a discharge with current-voltage profile shown in \fref{BiasVsCurrent}, with a peak discharge current of $100$ A, and corresponding current density of $5$ A/cm$^2$. The system employs argon gas at a pressure of 0.17 Pa, and an aluminum target about 5 cm in diameter. 
\begin{figure}[h!] 
\centering  {\includegraphics[height=8cm]{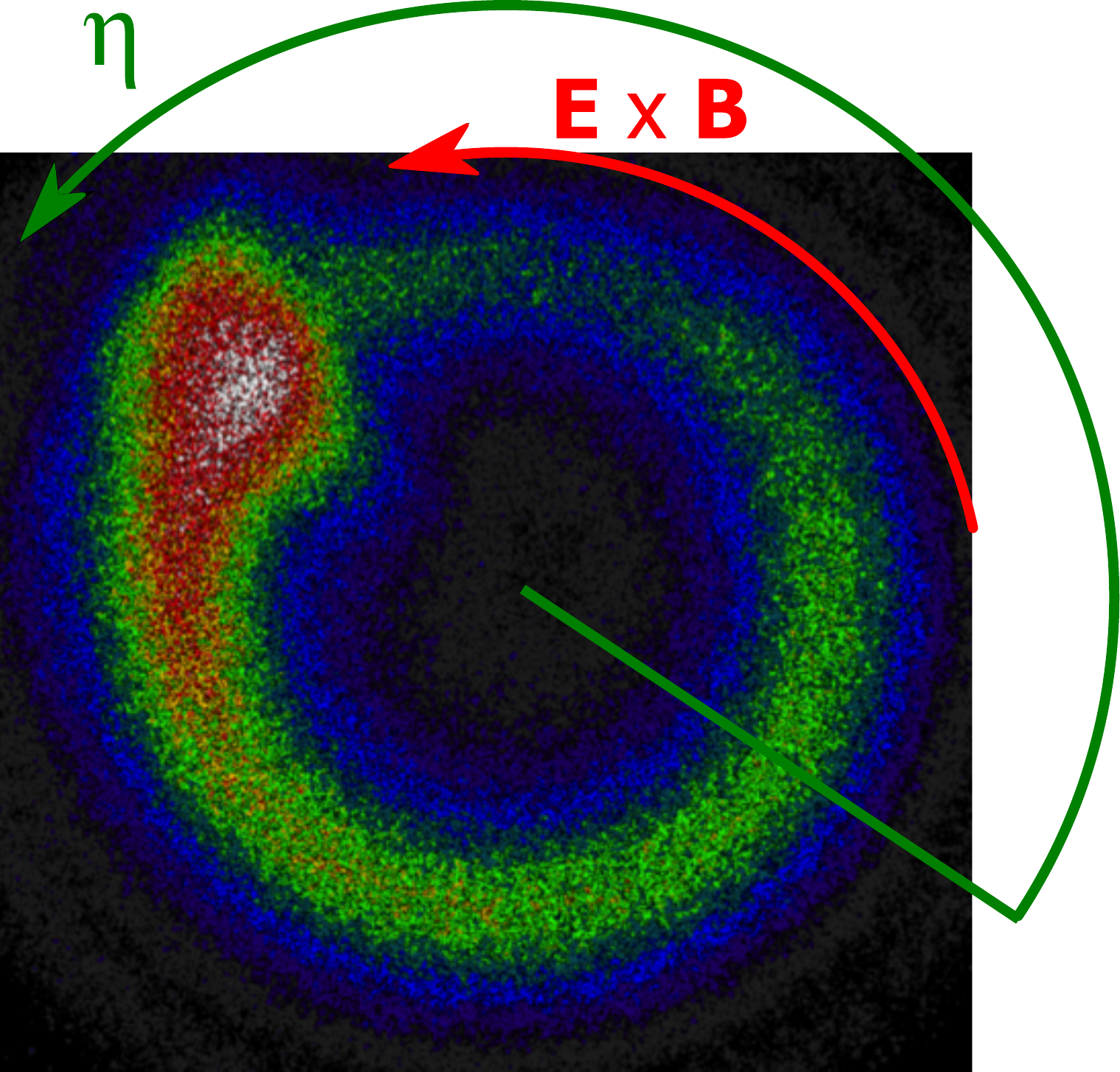}}\caption{Ionization region at approximately 100 A discharge current, as observed with an ICCD camera. The spoke rotates with constant rotation frequency of about $80$ kHz. Taken from Hecimovic \emph{et al.} \cite{2013-HecimovicPrivate}}\label{spoke}
\end{figure}
Pictures taken with an ICCD camera with $100$ ns acquisition time in the current plateau region, reveal the structure shown in \fref{spoke} rotating anti-clockwise with a constant rotation frequency of about $80$ kHz.\\
The aim of this simplified model is to qualitatively reproduce a hypothetical stable  configuration, in which a peak in the electron density remains unmodified and rotates with a constant angular velocity while it diffuses. No explanation for the formation of the peak is given here: the shape of the electron density is simply imposed, while the neutral density is calculated consistently.\\
We find that there exists a steady state configuration in a rotating frame with constant angular velocity, which is referred to as a ``rotating steady state", when the following mechanisms are taken into account: ground state ionization, electron loss, constant neutral refilling and the nonlinear term $\beta {n}^2_e n_{\rm n}$. 
\begin{figure}[h!] 
{\includegraphics[height=8cm]{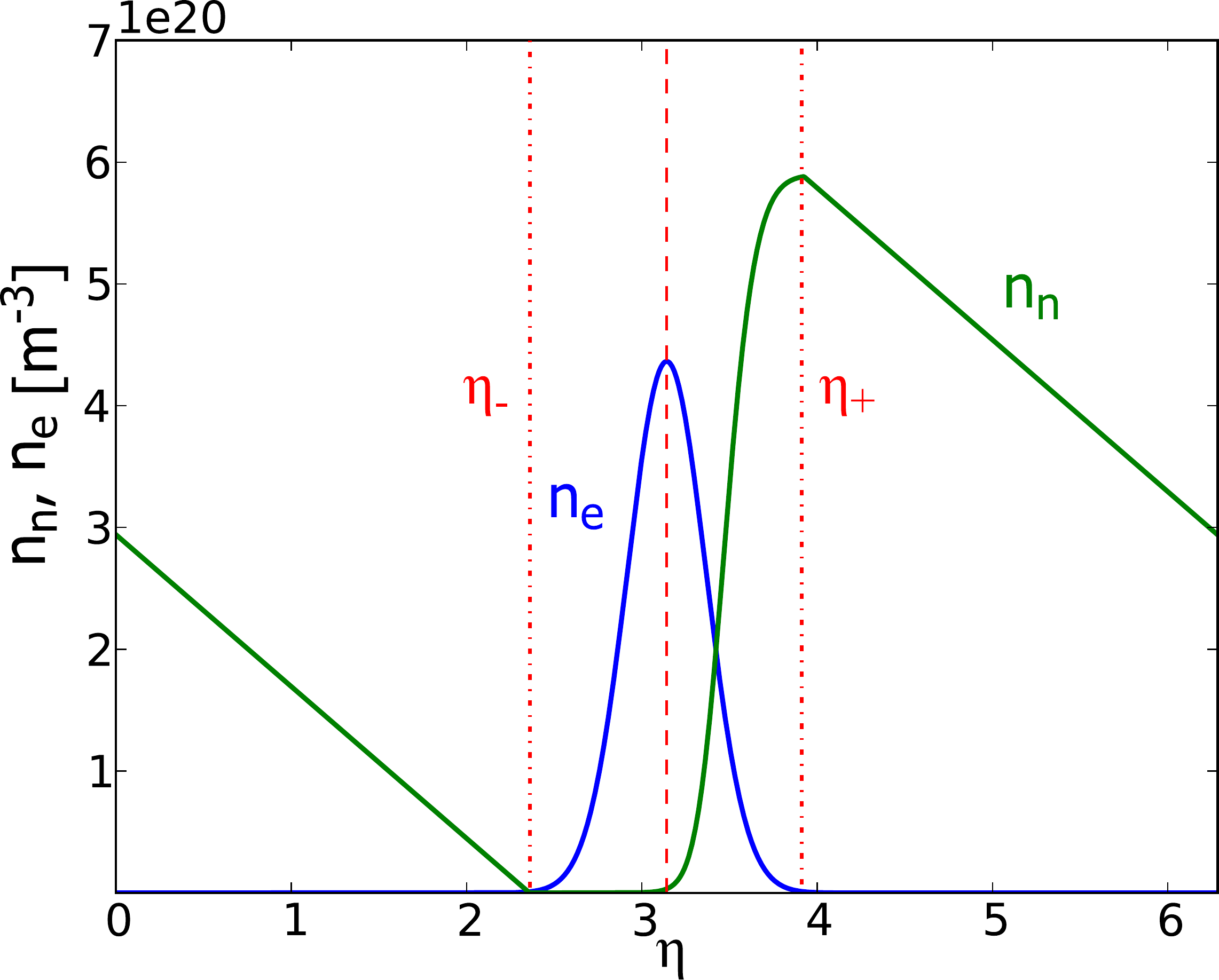}}\caption{Neutral (solid green) and electron (solid blue) densities vs $\eta = \theta - \Omega t$.}\label{nenn_stdcase}
\end{figure}
This result, even with the limitations underlined, suggests that the spoke experimentally observed can be sustained simply by the balance of the mentioned phenomena: ionization, loss and production of electrons and neutrals.\\
To specify a solution for the densities $n_{\rm n}(\eta)$ and $n_{\rm e}(\eta)$ for the set up considered, the physical parameters in equations \eref{neLF2} and \eref{nnLF2} need to be roughly estimated. The evaluation of the diffusion coefficient $D_\theta$, the ionization coefficient $k_{\rm ion}$, the electron loss rate $\nu_{\rm l}$ and the net neutral source term $R$ are addressed in \ref{Appendix}.\\
We have considered a region extending about $ L= \Delta z$ =  5 cm from the 2.5 cm radius magnetron surface. \\
The evaluation of the physical coefficients implies assumptions as to average plasma and neutral densities, and as to the ratios of metal to argon neutral and ion densities. 
But although in this sense the model is non self-consistent, it iterates on the imposed quantities until agreement with the densities obtained from the analytic solution is achieved. \ref{Appendix} describes the iteration procedure.
The degrees of ionization within the spoke that allow for simultaneous matching of the experimental discharge current and electron flux are $70 \%$ for argon and $75 \%$ for aluminum, corresponding to $\gamma_{\rm M} = 0.5$ and $\gamma_{\rm M+} = 0.56$. Such a high degree of ionization for both gas and metal is therefore necessary to match the experimental conditions, but one should keep in mind that it is restricted to the spoke region, where the electrons have a high density and all ionization phenomena take place.\\ 

The average magnetic field strength in the domain was $50$ mT and the electron temperature $3$ eV.  \\
The rotating steady state solution for the densities and their ratio are shown in \fref{nenn_stdcase}, with the passage of time causing movement from right to left. The electron density shows a peak in the spoke region, between $\eta_{-}$ and $\eta_+$, as imposed in \eref{gaussian} with amplitude calculated from \eref{nepeak}. Since quasi neutrality was assumed, all loss effects are driven by electrons: the electron loss rate $\nu_{\rm l}$ determines the overall loss rate and eventually the electron flux away from the target will equal the discharge current.\\
While the neutral density outside the spoke region increases due to diffusion of argon gas and sputtering of aluminum from the target at a constant rate, it drops sharply where the electron density increases as the neutrals are ionized. 
When the ratio of the densities $n_{\rm e}/n_{\rm n}$ approaches its maximum value, the neutral density reaches a minimum that remains constant till $\eta_{-}$. 
A comparison of $n_{\rm n}(\eta_-)$ and $n_{\rm n}(\eta_+)$ demonstrates an almost complete neutral depletion: the model does not allow one to state that almost $100 \%$ gas rarefaction is a necessary condition for the spoke existence and stability, as theorized in \cite{2013-BrenningLundin}, but shows that it is a consequence of the rotating steady state imposed.
 \\
Indeed this result shows that, in the critical conditions of the spoke region, it is possible to obtain almost full ionization of neutrals: $n_{\rm n}$ is primarily determined by the neutral refilling rate and the electron loss rate.
In the absence of a sufficient number of neutrals, the electrons cannot maintain the 
ionization rate at a level higher than the loss rate, so the electron density also drops.
When $n_{\rm e}$ drops to a very low value, $n_{\rm n}$ remains roughly constant.

\subsection{Solution sensitivity to physical parameters}
Given the empiricism of the derivation and estimation of the physical parameters, the sensitivity of the solution to the variation of the normalized $k_{\rm ion}$ and $\nu_{\rm l}$ is investigated. The aim of this section is to assess the choice of the parameters, by varying them in a large interval and arguing that the value chosen is the most sensible one. 

\begin{figure}[h!] 
{\includegraphics[height=8cm]{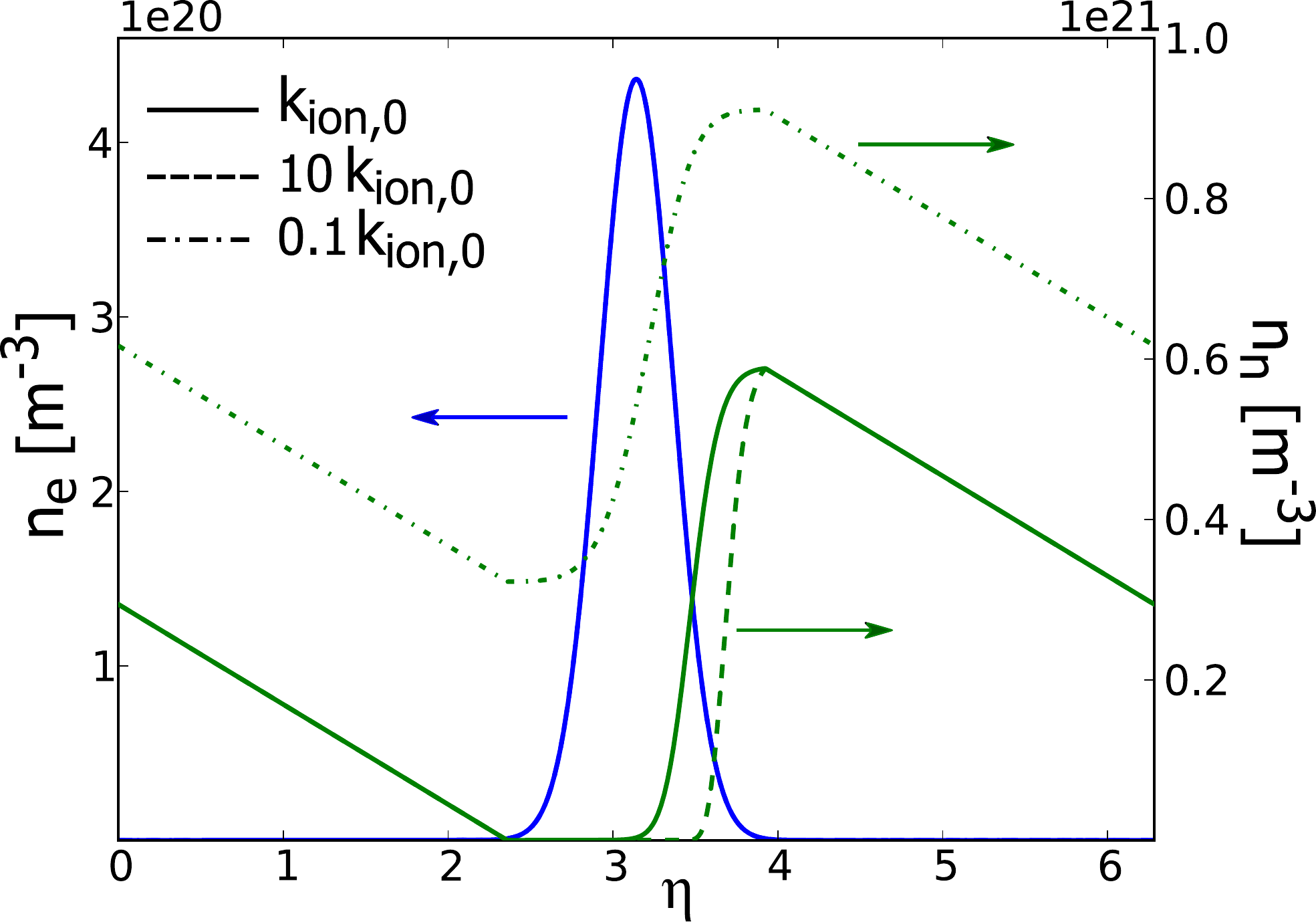}}
\caption{Sensitivity analysis of the solution varying the normalized ionization coefficient $k_{\rm ion}$ between $0.1\ k_{\rm ion,0}$ and $10\ k_{\rm ion,0}$. As expected the electron density is not influenced by $k_{\rm ion}$, while the neutral density is highly sensitive to its variation: if $k_{\rm ion}$ is too large the neutrals are ionized too quickly for the spoke to be sustained, on the other hand a small $k_{\rm ion}$ corresponds to a negligible depletion of the neutrals.}
\label{nenn_alpha}
\end{figure}

\paragraph{a) Varying the normalized ionization coefficient\\}
Figure \ref{nenn_alpha} shows the impact on the neutral solution of the variation of the non-dimensional coefficient $k_{\rm ion}$ between 0.1 $k_{\rm ion,0}$ and 10 $k_{\rm ion,0}$.  
Varying the ionization coefficient affects how quickly the neutrals are depleted within the electron density peak region: while a larger $k_{\rm ion}$ results in the neutral density reaching the minimum faster, a smaller $k_{\rm ion}$ radically changes the neutral density magnitude in the whole domain. An ionization coefficient one order of magnitude smaller would likely result in an insufficient neutral depletion for the spoke to appear. 
On the other hand, $k_{\rm ion}$ one order of magnitude larger results in a neutral depletion too rapid to be sustainable: the neutrals reach essentially full depletion before the peak in the electron density.

\begin{figure}[h!] 
{\includegraphics[height=8cm]{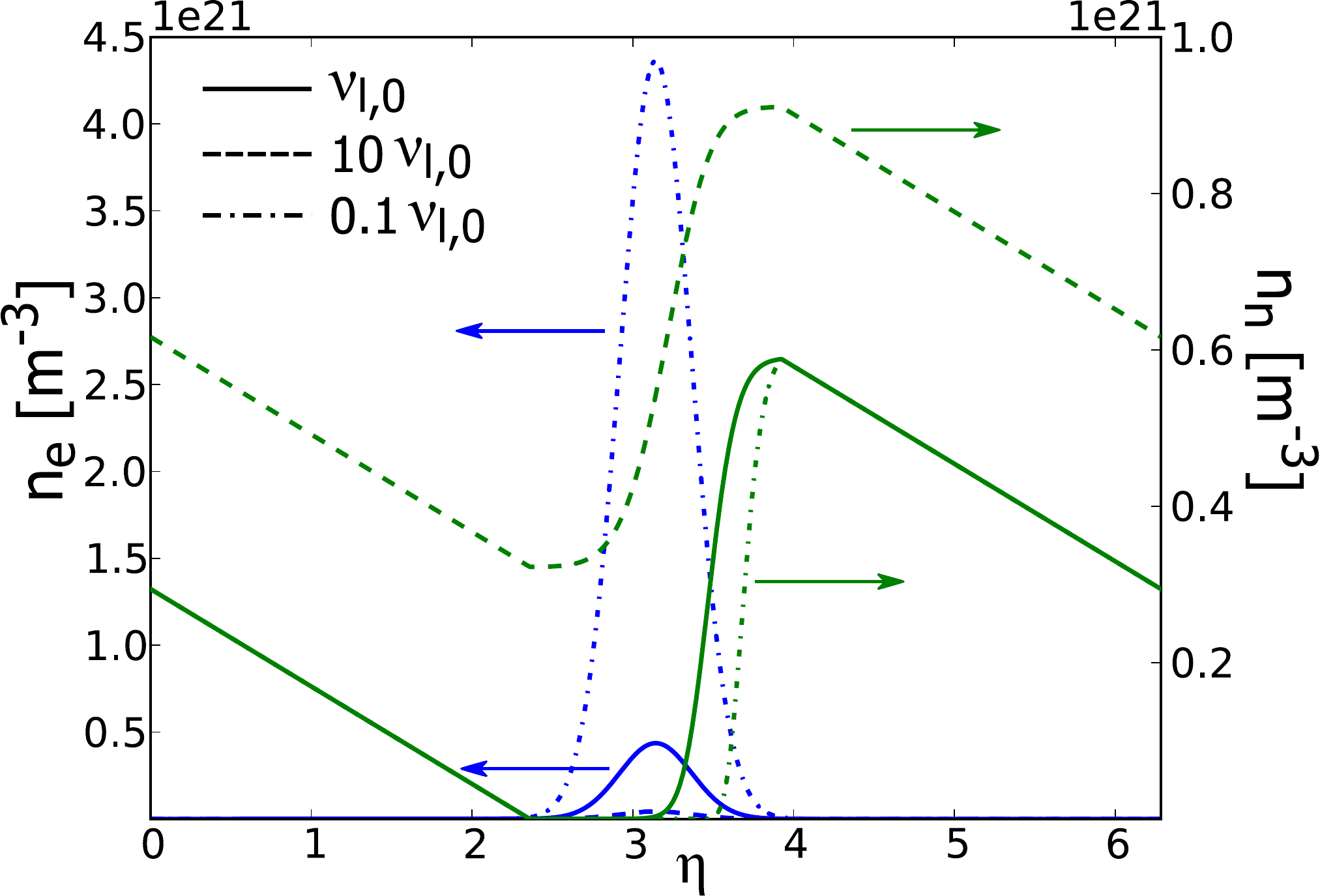}}
\caption{
Sensitivity analysis of the solution varying the normalized electron loss frequency $\nu_{\rm l}$ between $0.1\ \nu_{\rm l,0}$ and $10\ \nu_{\rm l,0}$. Both electron and neutral densities are sensitive to the electron loss frequency. In particular, the behavior of the neutral density is opposite to the case shown in figure \ref{nenn_alpha}: for small $\nu_{\rm l}$ the neutrals are depleted too quickly, while for large $\nu_{\rm l}$ the depletion is inconspicuous.
}\label{nenn_gamma}
\end{figure}

\paragraph{b) Varying the normalized electron loss frequency\\}
As expected, the solution for both densities shown in \fref{nenn_gamma} appears to be very sensitive to the variation of the electron loss frequency from the ionization region. 
When $\nu_{\rm l}$ is one order of magnitude smaller than estimated, the ratio of electron to neutral densities, $n_{\rm e}/n_{\rm n}$, becomes unrealistically large. 
Likewise, a larger $\nu_{\rm l}$ results in an electron density too small for the HiPIMS regime.\\

It can therefore be concluded that the order of magnitude of the estimated parameters is realistic for the description of the expected phenomenon.

\section{Conclusions}
When HiPIMS discharges are driven with sufficient power, well-defined emissivity peaks rotating in the $\textbf{E} \times \textbf{B}$ direction, and therefore breaking the azimuthal symmetry of the planar circular magnetron configuration, are observed. These regions of high density plasma are likely critical in the characterization of the overall discharge, therefore their understanding and modeling will shed light on the overall discharge behavior.
Starting from an advection diffusion reaction equation for electrons and two rate equations for working gas and metal neutrals a simplified model is devised. The terms retained consist of ionization, electron loss, constant neutral refilling and a nonlinear term. \\
When a Gaussian-like shape of the electron density $n_{\rm e}$ is imposed, the neutral density $n_{\rm n}$ and the magnitude of the electron density are analytically calculated in a frame co-moving with constant angular velocity with the high plasma density region.
The densities $n_{\rm e}$ and $n_{\rm n}$ are dependent on a series of physical parameters that are specified to reproduce the argon-aluminum discharge experimentally observed by A Hecimovic \emph{et al.}
A sensitivity analysis of the solution for the estimated parameters indicates that their order of magnitude is realistic to describe the phenomenon in a post hoc fashion.  \\
Despite the drastic simplification of the highly complex problem, the phenomenological model allows one to define a rotating steady state configuration of the electron and neutral densities by balancing of a reduced number of coexisting effects.
Since almost all neutrals are ionized by electrons in the high density region, the plasma density is primarily determined by the refilling rate and by the electron loss rate. The electron loss rate specifies the overall loss rate and therefore the confinement time. The electron density peak also is solved for consistently and results from the balance of electron diffusion in the azimuthal direction, electron loss in direction opposite to the target and neutral source rate. 
On the other hand, the neutral density is proportional to the amount of neutrals generated during a period of rotation of the spoke structure. \\
This model is intended as a merely qualitative descriptive tool, and represents only a first step toward devising a more sophisticated and eventually predictive model.


\appendix
\section{Determination of the physical parameters} \label{Appendix}

According to classical transport theory in presence of a magnetic field, assuming magnetized electrons ($\omega_{{\rm ce}}/\nu_{\rm T} \gg 1$), the diffusion coefficient in the azimuthal direction $D_\theta$ reads
\begin{equation}
	D_\theta \approx \frac{T_{\rm e}  \nu_{\rm T}}{m_{\rm e} \omega_{{\rm ce}}^2}  \frac{1}{r^2}, \label{Dperpclass}
\end{equation}
where $\omega_{{\rm ce}}$ is the gyro-frequency and $\nu_T$ is the effective collision frequency. The total collision frequency is given by the sum of the momentum transfer collision frequency with neutrals $\nu_{\rm m}$, and the bounce frequency along the magnetic field lines $\nu_{\rm bounce}$
\begin{equation*}
	\nu_{\rm T} = \nu_{\rm m} + \nu_{\rm bounce} \approx \frac{1}{2} \frac{v_{\rm th,e}}{L_{\rm ave}} + \left( k_{\rm m,Ar}(1-\gamma_M) + k_{\rm m,M} \gamma_M \right) \bar{n}_{\rm n}.
\end{equation*}
Electron transport has been observed to be anomalous in the direction perpendicular to the target in HiPIMS (e.g. \cite{2012-AndersNi,2008-LundinHelmersson,2004-BohlmarkHelmersson}), but for this rough estimate the classical formula is assumed to hold in the azimuthal direction.\\
The ionization coefficient $k_{\rm ion}$ is evaluated using the following expressions \cite{1990-FreundWetzel,2005-LiebermanLichtenberg} for the rate coefficients
\begin{eqnarray*}
	k_{\rm iz,Ar}(T_{\rm e}) &= \sigma_0 v_{\rm th,e} \left( 1 + \frac{2 T_{\rm e}}{\eps_{\rm iz}} \right) \exp({-\eps_{\rm iz}/T_{\rm e}}) ,\ {\rm  where } \ \sigma_0=\pi \left(\frac{q_{\rm e}}{4 \pi \eps_0 \eps_{\rm iz}}\right)^2, \\
	 k_{\rm iz,Al}(T_{\rm e}) &= 1.3467 \cdot 10^{-13} T_e^{0.3576} \exp(-6.7829/T_e).
\end{eqnarray*}

As concerns the excitation de-excitation term neglected in \eref{nnSteps}, a crude estimate of the rate coefficient $\left(- k_{\rm exc}  +  k_{\rm dexc}  n_{\rm Ar*}/n_{\rm Ar} \right)$ can be carried out using a simple steady state balance for Ar*
\begin{eqnarray*}
\frac{d n_{\rm Ar*}}{dt} = 0 = k_{\rm exc} n_e n_{\rm Ar} - k_{\rm iz, Ar*} n_e n_{\rm Ar*} - k_{\rm dexc} n_e n_{\rm Ar*} - k_{\rm p} n_e n_{\rm Ar*}\\
n_{\rm Ar*} = \frac{k_{\rm exc}}{k_{\rm dexc}+k_p+k_{\rm ion,Ar*}}
\end{eqnarray*}
that plugged into the neglected term in the first line of \eref{nnSteps} gives
\begin{equation*}
\left(- k_{\rm exc}  +  k_{\rm dexc}  \frac{n_{\rm Ar*}}{n_{\rm Ar}} \right) = k_{\rm exc} \left(-1 + \frac{k_{\rm dexc}}{k_{\rm dexc}+k_p+k_{\rm ion,Ar*}} \right) \approx -3 \cdot 10^{-16}, \quad {\rm m^3/s}
\end{equation*}
where the rate coefficients given in \cite{2011-RaaduAxnaes} have been used. This rate coefficient is therefore negligible compared to the ionization rate coefficient $k_{\rm ion} \approx 5 \cdot 10^{-14} \ m^3/s$, and the approximation made in \sref{syssetup} is legitimate.\\

The additional ionization rate $\beta$ in interpreted as a correction to the constant $k_{\rm ion}$, and therefore it is evaluated as $\beta = 0.1 \ k_{\rm ion}$. \\
The domain under consideration is in the $(r, \theta)$ plane, while it is the motion in the $z$ direction which is responsible for the loss of electrons. 
To a first approximation, one can estimate the magnitude of the velocity of the electrons leaving the system as
	\begin{equation*}
		|u_{z}| \approx \mu_{\perp,{\rm z}} E_{\rm z} + D_{\perp,{\rm z}} \frac{1}{\bar{n}_{\rm e}}\frac{\dd \bar{n}_{\rm e}}{\dd z},
	\end{equation*}
where $ \mu_{\perp,{\rm z}} \approx \frac{e}{m_e} \frac{\nu_{\rm T}}{\omega_{\rm c}^2} $ is the mobility across the field lines for magnetized electrons and $D_{\perp,{\rm z}}$ is the classical perpendicular diffusion coefficient as in \eref{Dperpclass}. 
The electron density $\bar{n}_{\rm e}(z)$ is averaged over the surface of the disk.  To estimate the loss rate of electrons $\nu_{\rm l}$, $|u_{z}|$ is averaged over $z$ between $z_0$ and $z_1$, which is assumed to account for the domain of interest (i.e. ionization region near the racetrack $L=z_1-z_0$)
\begin{equation*}
\fl\eqalign{
	\langle |u_{z}|\rangle_z \approx &\frac{1}{(z_1 - z_0)} \mu_{\perp,{\rm z}} \int_{z_0}^{z_1} \frac{d \phi}{d z} dz  + \frac{1}{(z_1 - z_0)}D_{\perp,{\rm z}}  \int_{z_0}^{z_1}  \frac{1}{\bar{n}_{\rm e}}\frac{d \bar{n}_{\rm e}}{d z} dz \\
				&  =\frac{1}{(z_1 - z_0)} \mu_{\perp,{\rm z}}[\phi(z_1)-\phi({z_0})] + \ \frac{D_{\perp,{\rm z}}}{(z_1 - z_0)} \ln \left(\frac{\bar{n}_{\rm e}(z_1)}{\bar{n}_{\rm e}(z_0)} \right )  \\ &\approx \frac{D_{\perp,{\rm z}}}{(z_1 - z_0)} \ln \left(\frac{\bar{n}_{\rm e}(z_1)}{\bar{n}_{\rm e}(z_0)} \right ) ,}
\end{equation*}
where the potential drop along $z$ has been neglected. Therefore, $\nu_{\rm l}$ reads
	\begin{equation*}
		\nu_{\rm l}\approx k_{ z}\frac{<|u_{\rm out}|>_z}{z_1-z_0} \approx k_{z} \frac{D_{\perp,{\rm z}}}{(z_1 - z_0)^2} \ln \left(\frac{\bar{n}_{\rm e}(z_1)}{\bar{n}_{\rm e}(z_0)} \right ).
	\end{equation*}
The coefficient $k_{z}$ is employed to adjust the cross field diffusion coefficient, that is known to be anomalous. $k_{z}$ is set to allow matching of the electron flux out of the domain with the experimental discharge current
\begin{equation*}
I_{\rm D} \approx q_{\rm e} \ \Gamma_{\rm e,z} \ S_{\rm RT}.
\end{equation*}
Finally, to evaluate the net source term for neutrals 
\begin{equation*}
R = \frac{\Gamma_{\rm Ar}-\Gamma_{\rm M}}{L} + \Gamma_{\rm e,z}[(1-\gamma_{\rm M+}) Y_{\rm sputt}+ \gamma_{\rm M+} Y_{\rm selfsputt}] \frac{S_{\rm RT}}{V},
\end{equation*}
we employ $Y_{\rm sputt} \approx 0.95$ for 500 eV Ar$^+$ on Al from \cite{1990-Ruzic} and $Y_{\rm selfsputt} \approx 1.1$ for 500 eV Al$^+$ on Al from \cite{1969-HaywardWolter}.
Following \cite{2011-RaaduAxnaes}, $\Gamma_{\rm M}$ results from Al diffusion away from the target: $\Gamma_{\rm M} = \Gamma_{\rm M,0} \exp(-L/\lambda_{\rm M,Ar})$, where $\Gamma_{\rm M,0} =  \bar{n}_{\rm M} v_{\rm th,M}/2 $ and $\lambda_{\rm M,Ar}$ is the Al mean free path in Ar. The argon flux from regions maintained at the pressure $p_{\rm 0,Ar}$ is estimated as: $\Gamma_{\rm Ar} = {n}_{\rm 0,Ar} (1-\bar{n}_{\rm Ar}/{n}_{\rm 0,Ar}) v_{\rm th,Ar}/2 $, where ${n}_{\rm 0,Ar} = p_{0,\rm Ar}/(k_{\rm B}T_{\rm Ar})$ from the ideal gas law.
Finally the electron flux $\Gamma_{\rm e,z}$ is evaluated from the experimental discharge current. To approximate the metal sputtering term as a constant, rather than retaining its dependence on the ion (therefore electron) density, makes the model non self-consistent but allows us to better reproduce the experimental operating conditions. 

\paragraph{Iteration procedure\\}

As mentioned in \ref{results}, the model is essentially non-self consistent: the ionization degrees $i_{\rm Ar}$ and $i_{\rm M}$ are imposed together with an average neutral density. From these data, the ratio $\gamma_{\rm M+}$ and an initial guess for the electron density is calculated. These in turn are used to evaluate the parameters $D_\theta$, $k_{\rm ion}$, $\nu_{\rm l}$ and $R$. 
Finally the coefficients are plugged into the analytic solution expressions to obtain the electron and neutral densities as a function of $\eta$, shown in \fref{nenn_stdcase}. The neutral density is then averaged and used as initial guess, the physical parameters are re-evaluated together with the solution. 
These steps are repeated until convergence.\\
Moreover, after convergence is reached, the combination of ionization degrees for argon and metal is chosen  to match simultaneously the discharge current $I_{\rm D}$ and electron flux $\Gamma_{\rm e,z}$ calculated as $1/2\ \bar{n}_{\rm e} u_{\rm B}$ and as $I_{\rm D}/(q_{\rm e} S_{\rm RT})$.


\ack
This work has been supported by the German Research Foundation (DFG) within the frame of the Collaborative Research Centre TRR 87 ``Pulsed high power plasmas for the synthesis of nanostructured functional layers" (SFB-TR 87). The authors gratefully acknowledge fruitful discussions with T de los Arcos, A Hecimovic and J Trieschmann.

\section*{References}

\bibliographystyle{iopart-num.bst}
\bibliography{biblio}
\newpage

\end{document}